\documentclass[10pt, a4paper, twocolumn, showpacs, titlepage, showkeys, aps]{revtex4-1}
\usepackage{times}
\usepackage{calligra}
\usepackage{hyperref}
\usepackage{amsmath}
\usepackage{amssymb}
\usepackage{graphicx}
\usepackage{subfigure}
\usepackage{booktabs}
\usepackage{rotating}
\usepackage{longtable}
\usepackage{bm}
\usepackage{color}
\usepackage{epstopdf}
\makeatletter

\newcommand{\Rmnum}[1]{\expandafter\@slowromancap\romannumeral #1@}
\makeatother

\begin{document}

\title{Tetrapartite entanglement features of W-Class state in uniform acceleration}
\author{Qian Dong$^{1}$}
\email{E-mail address: dldongqian@gmail.com (Q. Dong). }
\author{Ariadna J. Torres-Arenas$^{1}$}
\email{E-mail address: wia-k@hotmail.com (Ariadna J. Torres-Arenas). }
\author{Guo-Hua Sun$^{2}$}
\email{E-mail address: sunghdb@yahoo.com (G. H. Sun). }
\author{Shi-Hai Dong$^1$}
\email[Corresponding author:]{E-mail address: dongsh2@yahoo.com (S. H. Dong), Tel: 52-55-57296000 ext. 52522. }
\affiliation{$^1$ Laboratorio de Informaci\'{o}n Cu\'{a}ntica, CIDETEC, Instituto Polit\'{e}cnico Nacional, UPALM, CDMX 07700, Mexico}
\affiliation{$^2$ Catedr\'{a}tica CONACyT, Centro de Investigaci\'{o}n en Computaci\'{o}n, Instituto Polit\'{e}cnico Nacional, UPALM, CDMX 07738, Mexico}
\pacs{03. 67. -a, 03. 67. Mn, 03. 65. Ud, 04. 70. Dy}
\keywords{Tetrapartite, W-Class state, entanglement, Dirac field, noninertial frames}

\begin{abstract}
Using the single-mode approximation, we first calculate entanglement measures such as negativity ($1-3$ and $1-1$ tangles) and von Neumann entropy for a tetrapartite W-Class system in noninertial frame and then analyze the whole entanglement measures, the residual $\pi_{4}$ and geometric $\Pi_{4}$ average of tangles. Notice that the difference between $\pi_{4}$ and $\Pi_{4}$ is very small or disappears with the increasing accelerated observers. The entanglement properties are compared among the different cases from one accelerated observer to four accelerated observers. The results show that there still exists entanglement for the complete system even when acceleration $r$ tends to infinity. The degree of entanglement is disappeared for the $1-1$ tangle case when the acceleration $r > 0. 472473$. We reexamine the Unruh effect in noninertial frames. It is shown that the entanglement system in which only one qubit is accelerated is more robust than those entangled systems in which two or three or four qubits are accelerated. It is also found that the von Neumann entropy $S$ of the total system always increases with the increasing accelerated observers, but the $S_{\kappa\xi}$ and $S_{\kappa\zeta\delta}$ with two and three involved noninertial qubits first {\it increases} and then {\it decreases} with the acceleration parameter $r$, but they are equal to constants $1$ and $0. 811278$ respectively for zero involved noninertial qubit.

\end{abstract}
\maketitle

\section{Introduction}

One of the most studied notions of quantum correlations is the entanglement due to its important role in quantum information theory. The study of entanglement begins with Einstein, Podolsky and Rosen \cite{Einstein}, and Schr\"odinger \cite{Schrodinger1, Schrodinger2, Schrodinger3}  around 1930s. Now, entanglement is regarded as a key resource in quantum technology and it is often intertwined with quantum non-locality \cite{Werner, Horodecki, Guhne, Bell}. To quantify entanglement, a well justified and mathematically tractable measure is required. Negativity is one of the most common methods to quantify entanglement \cite{Peres, Zyczkowski} as well as whole entanglement \cite{Yazhou}. Also, another useful measurement is von Neumann entropy, relative entropy \cite{Vedral1, Vedral2, Vedral3}. Up to now, some works have been treated on bipartite systems except for a few multipartite systems \cite{Murao, Dur, Bennet3} since entanglement shared between two or multiple parties \cite{Horodecki, Modi, Alsing, Montero, Shamirzaie, Metwally} illustrates novel features. Collections of shared entangled qubits allow one to perform a number of quantum mechanical forms of communication, such as quantum dense coding and quantum teleportation \cite{Bennet1, Bennet2, Bouwmeester} since they play a significant role in efficient quantum communication \cite{Gisin, Terhal, Sen, yu16, yu17, yu18} and computational tasks \cite{Raussendorf, Briegel}.

In this work, we will investigate the tetrapartite entanglement of Dirac fields and consider the implementation of quantum information task between observers in uniform relative motion for a tetrapartite state, which is initially entangled in a W-Class state. This is because the quantum information in noninertial frame, which is a combination of general relativity, quantum field theory and quantum information theory, has been a focus of research topic in recent years. Its main aim is to incorporate relativistic effects to improve quantum-information tasks and to understand how such protocols will happen in curved space times. Since tripartite entangled state was worked out \cite{Alsing} and the Unruh effect was studied, most of the papers focus their investigation in two main states, i.e., Greenberger-Horne-Zeilinger (GHZ), W-state and other related states, but the W-state with less study due to the complexity of their calculations \cite{PRA_83_(2011)_012111, wang, ou, horn, seba, park16}. It should be pointed out that the computation of entanglement for the tripartite pure or mixed state in an accelerated frame is much more complicated because the density matrix cannot be written as the form of an X matrix. Nevertheless, we have recognized that the degree of entanglement for the W-Class state is more robust than that of the GHZ or relevant states {\color{red}\cite{dong18,peng10}}. This is an another reason why we attempt to carry out the W-Class entangled pure states even though its relevant calculations are rather complicated in comparison with other entangled states. Among the recent study on the Unruh effect in quantum information, it is found that in the fermionic case the degree to which entanglement is degraded depends on the election of Unruh modes. As done before, we also make use of the Rindler coordinates which define two disconnected regions I and II \cite{Takagi, Martin, Martin1}. For tetrapartite W-Class state, say Alice, Bob, Charlie and David, in this work we will consider all different cases from one accelerated observer to four accelerated observers and calculate their negativity, and whole entanglement $\pi_{4}$-tangle and $\Pi_{4}$-tangle but we restrict ourself to use the single mode approximation.

This work is organized as follows. In Section II we describe the tetrapartite entanglement of the W-Class for various cases which are from one accelerated observer to four accelerated observers. We obtain their density matrices and calculate their negativities ($1-1$ tangle and $1-3$ tangle) and whole entanglement measures. The Von Neumann entropy will be studied in Section III. Finally, some discussions and concluding remarks are given in Section IV.

\section{Tetrapartite entanglement from one to four accelerated observers}

A generalization for $N$ qubits of the W-Class entangled state which we are going to consider in this work has the form \cite{Dur2}:
\begin{equation}\label{w-class}
\left|W\right\rangle_{N}=\frac{1}{\sqrt{N}}\left|N-1, 1\right\rangle,
\end{equation}
where $\left|N-1, 1\right\rangle$ is a symmetric state involving a "1" and others $(N-1)$ "0"s. For the tetrapartite system $N=4$, the W-Class entangled state can be written as follows
\begin{equation}\label{w}
\begin{array}{l}
|W\rangle=\frac{1}{2}\left[|1_{\hat{A}}0_{\hat{B}}0_{\hat{C}}0_{\hat{D}}\right\rangle+\left|0_{\hat{A}}1_{\hat{B}}0_{\hat{C}}0_{\hat{D}}\right\rangle\\[2mm]
~~~~~~~~+\left|0_{\hat{A}}0_{\hat{B}}1_{\hat{C}}0_{\hat{D}}\right\rangle +\left|0_{\hat{A}}0_{\hat{B}}0_{\hat{C}}1_{\hat{D}}\right\rangle ],
\end{array}
\end{equation}where we use subscripts $A, B, C$ and $D$ to denote those observers and the Minkowski modes labeled with $M$ are omitted for observers $A, B, C$ and $D$.

For this entangled W-Class state in noninertial frame, it is conventional to use Rindler coordinates to describe this system. The Rindler coordinates describe a family of observers with uniform acceleration and divide Minkowski space-time into two inaccessible Regions I and II. The rightward accelerating observers are located in Region I and causally disconnected from their analogous counterparts in Region II \cite{Socolovsky, Mikio}.
We first give a brief review of the connection between the vacuum and excitation states in Minkowski coordinates and those in Rindler coordinates. Our setting consists of two observers: Alice and Bob. We first let Alice stay stationary, while Bob moves in uniform acceleration. Consider Bob to be accelerated uniformly in the $(t, z)$ plane. Rindler coordinates $(\tau, \xi)$ are appropriate for describing the viewpoint of an observer moving in uniform acceleration. Two different sets of the Rindler coordinates, which differ from each other by an overall change in sign, are necessary for covering Minkowski space. These sets of coordinates define two Rindler regions that are disconnected from each other \cite{Birrel and Davies, Alsing}
\begin{equation}\label{}
\begin{array}{l}
t=a^{-1}e^{a\xi}\sinh(a\tau), ~z=a^{-1}e^{a\xi}\cosh(a\tau), ~{\rm Region~ I}\\[2mm]
t=-a^{-1}e^{a\xi}\sinh(a\tau), ~z=-a^{-1}e^{a\xi}\cosh(a\tau), ~{\rm Region~ II}.
\end{array}
\end{equation}

A free Dirac field in $(3 + 1)$ dimensional Minkowski space satisfies the Dirac equation
\begin{equation}\label{}
i\gamma^{\mu}\partial_{\mu}\psi-m\psi=0,
\end{equation}
where $m$ is the particle mass, $\gamma^{\mu}$ are the Dirac gamma matrices, and $\psi$ is a spinor wave function, which is composed of the complete orthogonal set of fermion $\psi_{k}^{+}$, and antifermion $\psi_{k}^{-}$ modes and can be written as the following form
\begin{equation}\label{}
\psi=\int(a_{k}\psi_{k}^{+}+b_{k}^{\dagger}\psi_{k}^{-})dk,
\end{equation}where $a_{k}^{\dagger}(b_{k}^{\dagger})$ and $a_{k}(b_{k})$ are the creation and annihilation operators for fermions (antifermions) of the momentum $k$, respectively. They satisfy the anticommutation relation $\{a_{i}, a_{j}^{\dagger}\}=\{b_{i}, b_{j}^{\dagger}\}=\delta_{ij}$. The quantum field theory for a Rindler observer can be constructed by expanding the spinor field in terms of a complete set of fermion and antifermion modes in Regions I and II as
\begin{equation}\label{}
\psi=\int\sum_{\tau}(c_{k}^{\tau}\psi_{k}^{\tau+}+d_{k}^{\tau\dagger}\psi_{k}^{\tau-})dk, ~~~~\tau\in \{\rm I, \rm II\}.
\end{equation}

Similarly, $c_{k}^{\tau\dagger}(d_{k}^{\tau\dagger})$ and $c_{k}^{\tau}(d_{k}^{\tau})$ are the creation and annihilation operators for fermion (antifermions), respectively, acting on Region I~(II) for $\tau=\rm I~(\rm II)$ and satisfying similar anticommutation relation as above. The relation between creation and annihilation operators in Minkowski and Rindler space times can be found by using Bogoliubov transformation
\begin{equation}\label{}
a_{k}=\cos(r)\, c_{k}^{\rm I}-\sin(r)\, d_{-k}^{\rm II\dagger}, b_{k}=\cos(r)\, d_{k}^{\rm I}-\sin(r)\, c_{-k}^{\rm II\dagger},
\end{equation} where $\cos(r)=1/\sqrt{1+e^{-2\pi\omega_{k} c/a}}$ with $\omega_{k}=\sqrt{|\rm {\bf k}|^2+m^2}$ and $r$ is Bob's acceleration parameter with the range $r\in[0, \pi/4]$ for $a\in[0, \infty)$. It is seen from this equation and its adjoint that Bogoliubov transformation mixes a fermion in Region I and antifermions in Region II. As a result, it is assumed that the Minkowski particle vacuum state for mode $k$ based on Rindler Fock states is given by
\begin{equation}\label{0state}
|0_{k}\rangle_{M}=\sum_{n=0}^{1}A_{n}|n_{k}\rangle_{I}^{+}|n_{-k}\rangle_{II}^{-},
\end{equation}where the Rindler Region I or II Fock states carry a subscript I and II, respectively on the kets, but the Minkowski Fock states are indicated by the subscript $M$ on the kets. As what follows, we are only interested in using the {\it single mode approximation} \cite{Alsing, PRA_86_2012_012306, wang, arXiv1, qiangplb,qian18,arXiv1, annals2011}, i. e. , $w_{A,B,C,D}=w$ and also uniform acceleration $a_{A,B,C,D}=a$ ($a_{w,M}\approx a_{w,U}$ is considered to relate Minkowski and Unruh modes) for simplicity and we will drop all labels $(k,-k)$ on the states. Even though the single mode approximation is invalid for general states, however the approximation holds for a family of peaked Minkowski wave packets provided constraints imposed by an appropriate Fourier transform are satisfied \cite{bruschi10}.

Using the single mode approximation, Bob's vacuum state $|0_B\rangle$ and one-particle state $|1_B\rangle$ in Minkowski space are transformed into Rindler space. By applying the creation and annihilation operators to above equation (\ref{0state}) and using the normalization condition, we can obtain \cite{Alsing,PRA_86_2012_012306, wang, arXiv1, qiangplb,qian18,arXiv1, annals2011}
\begin{equation}\label{01}
\begin{array}{l}
|0\rangle_{M}=\cos (r) |0_\Rmnum{1} 0_{\Rmnum{2}}\rangle+\sin (r) |1_\Rmnum{1} 1_{\Rmnum{2}}\rangle, \\[2mm]
|1\rangle_{M} =|1_\Rmnum{1} 0_{\Rmnum{2}}\rangle,
\end{array}
\end{equation}
where $|n_{B_\Rmnum{1}}\rangle$ and $|n_{B_{\Rmnum{2}}}\rangle$ ($n=0, 1$) are the mode decomposition of $|n_B\rangle$ into two causally disconnected Regions I and II in Rindler space. It should be pointed out that Bruschi {\it et al. } discussed the Unruh effect {\it beyond} the {\it single mode approximation} \cite{bruschi10}, in which two complex numbers $q_{R}$ and $q_{L}$ (the subindexes $L$ and $R$ corresponding to the Left and Right regions in Rindler diagram, i. e. Regions I and II) are used to construct the one-particle state, i. e. , $|1\rangle=q_{R}|1_{R}0_{L}\rangle+q_{L}|0_{R}1_{L}\rangle$. However, in the present case for single mode approximation one has $q_{R}=1, q_{L}=0$ to satisfy the normalization condition $|q_{R}|^2+|q_{L}|^2=1$. It is also worth noting that a Minkowski mode that defines the Minkowski vacuum is related to a highly nonmonochromatic Rindler mode rather than a single mode with the same frequency (see Refs. \cite{Martin,Martin1,bruschi10} for details). Other relevant contributions \cite{eduardo12,bruschi13,bruschiclass,eduardo13, mann13} have also been made.

Since the moving observers are confined to Region I, we have to trace out the part of the antiparticle state in Region II. Let us apply Eq. (\ref{01}) to the $\left|W\right\rangle$ state (\ref{w}). We study this entanglement system in four different cases. First, we study the case when David is accelerated,
\begin{equation}
\begin{array}{l}
\left|W_{D}\right\rangle=\displaystyle\frac{1}{2}\Big[\sin  r (\left|0_{\hat{A}} 0_{\hat{B}} 1_{\hat{C}} 1_{\hat{\text{DI}}}\right\rangle + \left|0_{\hat{A}} 1_{\hat{B}} 0_{\hat{C}} 1_{\hat{\text{DI}}}\right\rangle\\[2mm]
~~~~~~~~~+\left|1_{\hat{A}} 0_{\hat{B}}0_{\hat{C}}1_{\hat{\text{DI}}}\right\rangle)+\cos  r (\left|0_{\hat{A}}0_{\hat{B}}1_{\hat{C}}0_{\hat{\text{DI}}}\right\rangle \\[2mm]
~~~~~~~~~+ \left|0_{\hat{A}}1_{\hat{B}}0_{\hat{C}}0_{\hat{\text{DI}}}\right\rangle +\left|1_{\hat{A}}0_{\hat{B}}0_{\hat{C}}0_{\hat{\text{DI}}}\right\rangle )+\left|0_{\hat{A}}0_{\hat{B}}0_{\hat{C}}1_{\hat{\text{DI}}}\right\rangle\Big].
\end{array}
\end{equation}

Second, we consider the case when Charlie and David are accelerated,
\begin{equation}
\begin{array}{l}
\left|W_{CD}\right\rangle=\displaystyle\frac{1}{2}\Big[ \sin ^2 r (\left|0_{\hat{A}}1_{\hat{B}}1_{\hat{\text{CI}}}1_{\hat{\text{DI}}}\right\rangle +\left|1_{\hat{A}}0_{\hat{B}}1_{\hat{\text{CI}}} 1_{\hat{\text{DI}}}\right\rangle)\\[2mm]
~~~~~~~~~~~~~~~+ \cos ^2 r (\left|0_{\hat{A}}1_{\hat{B}}0_{\hat{\text{CI}}}0_{\hat{\text{DI}}}\right\rangle + \left|1_{\hat{A}}0_{\hat{B}}0_{\hat{\text{CI}}}0_{\hat{\text{DI}}}\right\rangle)\\[2mm]
~~~~~~~~~~~~~~~+\cos r \sin r (\left|0_{\hat{A}}1_{\hat{B}}0_{\hat{\text{CI}}}1_{\hat{\text{DI}}}\right\rangle+\left|0_{\hat{A}}1_{\hat{B}}1_{\hat{\text{CI}}}0_{\hat{\text{DI}}}\right\rangle\\
~~~~~~~~~~~~~~~+ \left|1_{\hat{A}}0_{\hat{B}}0_{\hat{\text{CI}}}1_{\hat{\text{DI}}}\right\rangle+\left|1_{\hat{A}}0_{\hat{B}}1_{\hat{\text{CI}}}0_{\hat{\text{DI}}}\right\rangle)\\[2mm]
~~~~~~~~~~~~~~~+ \sin r (\left|0_{\hat{A}}0_{\hat{B}}1_{\hat{\text{CI}}}1_{\hat{\text{DI}}}\right\rangle+ \left|0_{\hat{A}}0_{\hat{B}}1_{\hat{\text{CI}}}1_{\hat{\text{DI}}}\right\rangle)\\
~~~~~~~~~~~~~~~+ \cos r (\left|0_{\hat{A}}0_{\hat{B}}0_{\hat{\text{CI}}}1_{\hat{\text{DI}}}\right\rangle+ \left|0_{\hat{A}}0_{\hat{B}}1_{\hat{\text{CI}}}0_{\hat{\text{DI}}}\right\rangle)\Big].
\end{array}
\end{equation}

Third, we consider the case when Bob, Charlie and David are accelerated,
\begin{widetext}
\begin{equation}
\begin{array}{l}
\left|W_{BCD}\right\rangle=\displaystyle\frac{1}{2}\Big(\sin ^3  r \left|1_{\hat{A}} 1_{\hat{\text{BI}}} 1_{\hat{\text{CI}}} 1_{\hat{\text{DI}}}\right\rangle +
\cos ^3  r \left|1_{\hat{A}} 0_{\hat{\text{BI}}} 0_{\hat{\text{CI}}} 0_{\hat{\text{DI}}}\right\rangle +
\cos ^2  r \sin  r \left|1_{\hat{A}} 0_{\hat{\text{BI}}} 0_{\hat{\text{CI}}} 1_{\hat{\text{DI}}}\right\rangle +\cos ^2  r \sin  r \left|1_{\hat{A}} 0_{\hat{\text{BI}}} 1_{\hat{\text{CI}}} 0_{\hat{\text{DI}}}\right\rangle\\[2mm]
~~~~~~~~~~~~~~~+\cos  r \sin ^2  r \left|1_{\hat{A}} 0_{\hat{\text{BI}}} 1_{\hat{\text{CI}}} 1_{\hat{\text{DI}}}\right\rangle +
\sin  r \cos ^2  r \left|1_{\hat{A}} 1_{\hat{\text{BI}}} 0_{\hat{\text{CI}}} 0_{\hat{\text{DI}}}\right\rangle +\sin ^2  r \cos  r \left|1_{\hat{A}} 1_{\hat{\text{BI}}} 0_{\hat{\text{CI}}} 1_{\hat{\text{DI}}}\right\rangle +
\sin ^2  r \cos  r \left|1_{\hat{A}} 1_{\hat{\text{BI}}} 1_{\hat{\text{CI}}} 0_{\hat{\text{DI}}}\right\rangle\\[2mm]
~~~~~~~~~~~~~~~+\sin ^2  r \left|0_{\hat{A}} 1_{\hat{\text{BI}}} 1_{\hat{\text{CI}}} 1_{\hat{\text{DI}}}\right\rangle +\cos ^2  r \left|0_{\hat{A}} 0_{\hat{\text{BI}}} 0_{\hat{\text{CI}}} 1_{\hat{\text{DI}}}\right\rangle+
\cos  r \sin  r \left|0_{\hat{A}} 0_{\hat{\text{BI}}} 1_{\hat{\text{CI}}} 1_{\hat{\text{DI}}}\right\rangle +
\sin  r \cos  r \left|0_{\hat{A}} 1_{\hat{\text{BI}}} 0_{\hat{\text{CI}}} 1_{\hat{\text{DI}}}\right\rangle\\[2mm]
~~~~~~~~~~~~~~~+\sin  r \cos  r \left|0_{\hat{A}} 1_{\hat{\text{BI}}} 1_{\hat{\text{CI}}} 0_{\hat{\text{DI}}}\right\rangle +
\sin ^2  r \left|0_{\hat{A}} 1_{\hat{\text{BI}}} 1_{\hat{\text{CI}}} 1_{\hat{\text{DI}}}\right\rangle +
\cos ^2  r \left|0_{\hat{A}} 1_{\hat{\text{BI}}} 0_{\hat{\text{CI}}} 0_{\hat{\text{DI}}}\right\rangle+\sin ^2  r \left|0_{\hat{A}} 1_{\hat{\text{BI}}} 1_{\hat{\text{CI}}} 1_{\hat{\text{DI}}}\right\rangle \\[2mm]
~~~~~~~~~~~~~~~+\cos ^2  r \left|0_{\hat{A}} 0_{\hat{\text{BI}}} 1_{\hat{\text{CI}}} 0_{\hat{\text{DI}}}\right\rangle +
\cos  r \sin  r \left|0_{\hat{A}} 0_{\hat{\text{BI}}} 1_{\hat{\text{CI}}} 1_{\hat{\text{DI}}}\right\rangle+\cos  r \sin  r\left|0_{\hat{A}} 1_{\hat{\text{BI}}} 0_{\hat{\text{CI}}} 1_{\hat{\text{DI}}}\right\rangle +
\sin  r \cos  r \left|0_{\hat{A}} 1_{\hat{\text{BI}}} 1_{\hat{\text{CI}}} 0_{\hat{\text{DI}}}\right\rangle\Big).
\end{array}
\end{equation}
\end{widetext}
Fourth, we will study the case when Alice, Bob, Charlie and David are accelerated,
\begin{widetext}
\begin{equation}
\begin{array}{l}
\left|W_{ABCD}\right\rangle=\displaystyle\frac{1}{2}\Big(\cos ^3  r \left|0_{\hat{\text{AI}}}0_{\hat{\text{BI}}}0_{\hat{\text{CI}}}1_{\hat{\text{DI}}}\right\rangle +
\cos ^3  r \left|0_{\hat{\text{AI}}}0_{\hat{\text{BI}}}1_{\hat{\text{CI}}}0_{\hat{\text{DI}}}\right\rangle +
\cos ^2 \sin  r \left|0_{\hat{\text{AI}}}0_{\hat{\text{BI}}}1_{\hat{\text{CI}}}1_{\hat{\text{DI}}}\right\rangle+\cos ^2  r \sin  r \left|0_{\hat{\text{AI}}}1_{\hat{\text{BI}}}0_{\hat{\text{CI}}}1_{\hat{\text{DI}}}\right\rangle \\[2mm]
~~~~~~~~~~~~~~+\cos ^2  r \sin  r \left|0_{\hat{\text{AI}}}1_{\hat{\text{BI}}}1_{\hat{\text{CI}}}0_{\hat{\text{DI}}}\right\rangle +
\cos ^2  r \sin  r \left|0_{\hat{\text{AI}}}1_{\hat{\text{BI}}}1_{\hat{\text{CI}}}0_{\hat{\text{DI}}}\right\rangle +\cos  r \sin ^2  r \left|0_{\hat{\text{AI}}}1_{\hat{\text{BI}}}1_{\hat{\text{CI}}}1_{\hat{\text{DI}}}\right\rangle +
\cos  r \sin ^2  r \left|0_{\hat{\text{AI}}}1_{\hat{\text{BI}}}1_{\hat{\text{CI}}}1_{\hat{\text{DI}}}\right\rangle \\[2mm]
~~~~~~~~~~~~~~+\cos  r \sin ^2  r \left|0_{\hat{\text{AI}}}1_{\hat{\text{BI}}}1_{\hat{\text{CI}}}1_{\hat{\text{DI}}}\right\rangle +\sin  r \cos^2  r \left|1_{\hat{\text{AI}}}0_{\hat{\text{BI}}}1_{\hat{\text{CI}}}0_{\hat{\text{DI}}}\right\rangle +
\cos^2  r \sin  r \left|1_{\hat{\text{AI}}}0_{\hat{\text{BI}}}1_{\hat{\text{CI}}}0_{\hat{\text{DI}}}\right\rangle +
\sin^2  r \cos  r \left|1_{\hat{\text{AI}}}0_{\hat{\text{BI}}}1_{\hat{\text{CI}}}1_{\hat{\text{DI}}}\right\rangle \\[2mm]
~~~~~~~~~~~~~~+\sin^2  r \cos  r \left|1_{\hat{\text{AI}}}0_{\hat{\text{BI}}}1_{\hat{\text{CI}}}1_{\hat{\text{DI}}}\right\rangle +
\cos  r \sin^2  r \left|1_{\hat{\text{AI}}}0_{\hat{\text{BI}}}1_{\hat{\text{CI}}}1_{\hat{\text{DI}}}\right\rangle +
\sin  r \cos^2  r \left|1_{\hat{\text{AI}}}1_{\hat{\text{BI}}}0_{\hat{\text{CI}}}0_{\hat{\text{DI}}}\right\rangle +\sin  r \cos^2  r \left|1_{\hat{\text{AI}}}1_{\hat{\text{BI}}}0_{\hat{\text{CI}}}0_{\hat{\text{DI}}}\right\rangle \\[2mm]
~~~~~~~~~~~~~+\sin^2  r \cos  r \left|1_{\hat{\text{AI}}}1_{\hat{\text{BI}}}0_{\hat{\text{CI}}}1_{\hat{\text{DI}}}\right\rangle +
\sin^2  r \cos  r \left|1_{\hat{\text{AI}}}1_{\hat{\text{BI}}}0_{\hat{\text{CI}}}1_{\hat{\text{DI}}}\right\rangle +\sin^2  r \cos  r \left|1_{\hat{\text{AI}}}1_{\hat{\text{BI}}}0_{\hat{\text{CI}}}1_{\hat{\text{DI}}}\right\rangle +
\sin^2  r \cos  r \left|1_{\hat{\text{AI}}}1_{\hat{\text{BI}}}1_{\hat{\text{CI}}}0_{\hat{\text{DI}}}\right\rangle \\[2mm]
~~~~~~~~~~~~~+\sin^2  r \cos  r \left|1_{\hat{\text{AI}}}1_{\hat{\text{BI}}}1_{\hat{\text{CI}}}0_{\hat{\text{DI}}}\right\rangle +\cos ^3  r \left|1_{\hat{\text{AI}}}0_{\hat{\text{BI}}}0_{\hat{\text{CI}}}0_{\hat{\text{DI}}}\right\rangle +
\sin  r \cos ^2  r \left|1_{\hat{\text{AI}}}0_{\hat{\text{BI}}}0_{\hat{\text{CI}}}1_{\hat{\text{DI}}}\right\rangle +
\cos^2  r \sin  r \left|1_{\hat{\text{AI}}}0_{\hat{\text{BI}}}0_{\hat{\text{CI}}}1_{\hat{\text{DI}}}\right\rangle\\[2mm]
~~~~~~~~~~~~~~+\cos ^2  r \sin  r \left|0_{\hat{\text{AI}}}0_{\hat{\text{BI}}}1_{\hat{\text{CI}}}1_{\hat{\text{DI}}}\right\rangle +
\cos ^3  r \left|0_{\hat{\text{AI}}}1_{\hat{\text{BI}}}0_{\hat{\text{CI}}}0_{\hat{\text{DI}}}\right\rangle +
\cos ^2  r \sin  r \left|0_{\hat{\text{AI}}}1_{\hat{\text{BI}}}0_{\hat{\text{CI}}}1_{\hat{\text{DI}}}\right\rangle +\sin^2  r \cos  r \left|1_{\hat{\text{AI}}}1_{\hat{\text{BI}}}1_{\hat{\text{CI}}}0_{\hat{\text{DI}}}\right\rangle\\[2mm]
~~~~~~~~~~~~~~+\sin^3  r \left|1_{\hat{\text{AI}}}1_{\hat{\text{BI}}}1_{\hat{\text{CI}}}1_{\hat{\text{DI}}}\right\rangle +
\sin^3  r \left|1_{\hat{\text{AI}}}1_{\hat{\text{BI}}}1_{\hat{\text{CI}}}1_{\hat{\text{DI}}}\right\rangle
+\sin^3  r \left|1_{\hat{\text{AI}}}1_{\hat{\text{BI}}}1_{\hat{\text{CI}}}1_{\hat{\text{DI}}}\right\rangle +
\sin^3  r \left|1_{\hat{\text{AI}}}1_{\hat{\text{BI}}}1_{\hat{\text{CI}}}1_{\hat{\text{DI}}}\right\rangle\Big).
\end{array}
\end{equation}
\end{widetext}

As what follows, we will study these different cases for their negativities and von Neumann entropy to show their entanglement properties which are related to the uniform acceleration $r$.

\subsection{Negativity}

As a quantitative entanglement measure, the negativity has been computed for many entangled systems. Negativity quantifies the entanglement in a state as the degree to see whether the entangled system is still entangled or not. An entangled system $\rho$ is entangled if there exists at least one negative eigenvalue for the partial transpose of the corresponding density matrix. The negativity for a tetrapartite state is defined as \cite{Yazhou}
\begin{equation}\label{}
N_{\kappa (\xi \o \zeta)}=||\rho_{\kappa(\xi \o \zeta)}^{T_{\kappa}}||-1, ~~~~N_{\kappa \xi}=||\rho_{\kappa \xi}^{T_{\kappa}}||-1,
\end{equation}which describe the entanglements $1-3$ tangle and $1-1$ tangle, respectively. The notations $||\rho_{\kappa \xi \o \zeta}^{T_{\kappa}}||$ and $||\rho_{\kappa \xi}^{T_{\kappa}}||$ are the trace-norm of each partial transpose matrix.

Alternatively, since $||O||={\rm tr} \sqrt{O^{\dagger} O}$ for any Hermitian operator $O$ \cite{Williams}, one can write
\begin{equation}\label{neg}
||M||-1=2\sum_{i=1}^{N}|\lambda_{M}^{(-)}|^{i},
\end{equation}
where $\lambda_{M}^{(-)}$ are the negative eigenvalues of the matrix $M$.

After obtaining the density matrix of each system and tracing out the antiparticle in Region II, we proceed to find the negative eigenvalues of each density matrix in order to solve equation (\ref{neg}). This will make us find negativities ($1-3$ tangle) for $N_{A(BCD)}, N_{B(ACD)}, N_{C(ABD)}, N_{D(ABC)}$ by varying the quantities of accelerated qubits 1, 2, 3 or all. Analytical expressions of the negativity are not written out due to their complications, while we illustrate them in FIG. 1. Considering the symmetry of this entangled system, we have $N_{A(BCD_{I})}=N_{B(ACD_{I})}=N_{C(ABD_{I})}$, $N_{C_I(ABD_I)}=N_{D_I(ABC_I)}$, $N_{B_I(AC_ID_I)}=N_{C_I(AB_ID_I)}=N_{D_I(AB_IC_I)}$ and $N_{A_I(B_IC_ID_I)}=N_{B_I(A_IC_ID_I)}=N_{C_I(A_IB_ID_I)}$. We notice that the entanglement degree decreases along with the increasing accelerated observers. This means that the entanglement system in which only one qubit is accelerated is more robust than those entangled systems in which two or three or four qubits are accelerated. It should be recognized that the degree of entanglement for each system has never been disappeared even in the infinite acceleration.

\begin{figure}[htbp]
\includegraphics[width=8cm]{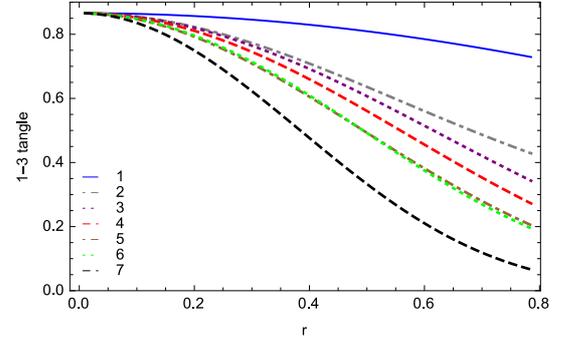}\label{1-3 tangle}\hfil%
\caption{\label{1-3 tangle} (Color online) The $1-3$ tangle negativities as a function of the acceleration parameter $r$.
The blue solid line "1" corresponds to $N_{A(BCD_{I})}=N_{B(ACD_{I})}=N_{C(ABD_{I})}$, the gray dotdashed and purple dotted lines "2" and $3$ correspond to $N_{A(BC_{I}D_I)}$ and $N_{D_I(ABC)}$, respectively. The red dashed
line "4" corresponds to $N_{C_I(ABD_I)}=N_{D_I(ABC_I)}$, the brown dotdashed line "5" corresponds to $N_{A(B_I C_I D_I)}$, the green dotted line "6" corresponds to $N_{B_I(AC_ID_I)}=N_{C_I(AB_ID_I)}=N_{D_I(AB_IC_I)}$ and the black dashed line "7" corresponds to $N_{A_I(B_IC_ID_I)}=N_{B_I(A_IC_ID_I)}=N_{C_I(A_IB_ID_I)}$. }

\end{figure}

On the other hand, it is also important to find the $1-1$ tangle which is required to calculate the whole entanglement measures. With the similar process to the $1-3$ tangle case, we also trace out the necessary qubits and generate bipartite subsystems with all possible combinations of all qubits. Importantly, we might use the symmetry between the negativity computations for each pair of qubits to get those commutative negativities. The corresponding results are plotted in FIG. 2. There are 24 analytical results of the $1-1$ tangle, which have the following possible values,
\begin{equation}
\begin{array}{l}
N_{\kappa \xi}=\displaystyle\frac{1}{2} (\sqrt{2}-1)=0. 2071, \\[2mm]
N_{\kappa_{I} \xi}=\displaystyle\frac{1}{16} \Big[-2 \cos (2 r)-6\\[2mm]
~~~~~~~+\sqrt{2} \sqrt{28 \cos (2 r) +9 \cos (4 r)+27}\Big], \\[2mm]
N_{\kappa_{I} \xi_{I}}=\displaystyle\frac{1}{8} \Big[2 \cos (2 r)- \cos (4 r)-5\\[2mm]
~~~~~~~+2 \sqrt{5 \cos (4 r) -4 ( \cos (2 r) )+7}\Big],
\end{array}
\end{equation}
where $N_{\kappa \xi} > N_{\kappa_{I} \xi} > N_{\kappa_{I} \xi_{I}}$ and the $N_{\kappa \xi}$, $N_{\kappa_{I} \xi}$ and $N_{\kappa_{I} \xi_{I}}$ with $\kappa,\xi\in(A, B, C, D)$, are all possible subsystem combinations with two inertial qubits, one inertial qubit and without any inertial qubit. It is interesting to see that the degree of entanglement will be disappeared for the acceleration parameter $r>0. 472473$ in the case of the four qubits being accelerated simultaneously.

\begin{figure}[htbp]
\includegraphics[width=9cm]{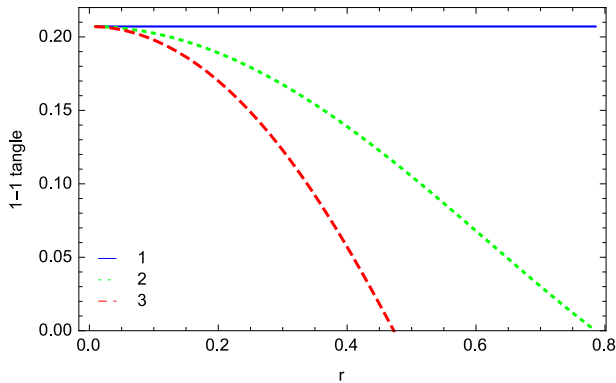}
\caption{\label{1-1 tangle}(Color online) The $1-1$ tangle negativities as a function of the acceleration parameter $r$. The blue solid line "1" corresponds to $N_{A(B)}=N_{A(C)}=N_{B(C)}$, the green dotted line "2" corresponds to $N_{A(B_I)}=N_{A(C_I)}=N_{A(D_I)}=N_{B(C_I)}=N_{B(D_I)}=N_{C(D_I)}$ and the red dashed line "3" corresponds to $N_{A_I(B_I)}=N_{A_I(C_I)}=N_{A_I(D_I)}=N_{B_I(C_I)}=N_{B_I(D_I)}=N_{C_I(D_I)}$. }
\end{figure}

\begin{figure}[htbp]{\includegraphics[width=9cm]{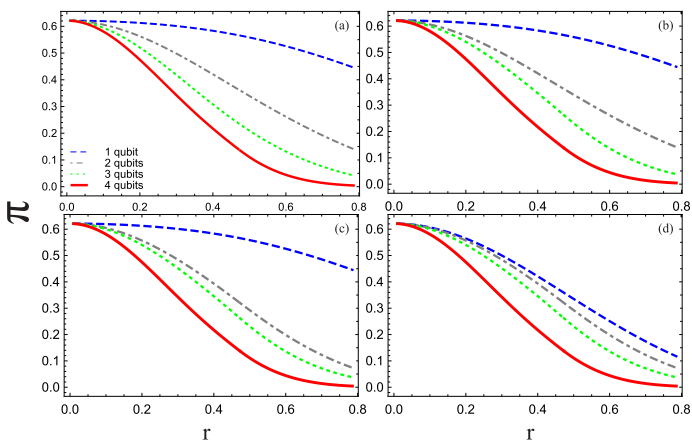}}\label{Residual entanglement}
\caption{\label{Residual entanglement} (Color online) Residual entanglement of Alice $\pi_A$, Bob $\pi_B$, Charlie $\pi_C$ and David $\pi_D$  as a function of the acceleration parameter $r$ illustrated in (a), (b), (c) and (d), respectively.}
\end{figure}

\subsection{Whole entanglement measures }

Another quantification of multipartite entanglement is the residual tangle $\pi_4$. The residual tangle which measures entanglement among the four components can be calculated by the following form \cite{Oliveira} (see FIG. 3)
\begin{eqnarray}\label{eq24-27}
\pi_{\kappa}=N_{\kappa(\xi \o \zeta)}^{2}-N_{\kappa \xi}^{2}-N_{\kappa \o}^{2}-N_{\kappa \zeta}^2,\\
\pi_{\xi}=N_{\xi(\kappa \o \zeta)}^{2}-N_{\xi\kappa}^{2}-N_{\xi \o}^{2}-N_{\xi \zeta}^2,\\
\pi_{\o}=N_{\o(\kappa \xi \zeta)}^{2}-N_{\o\kappa}^{2}-N_{\o\xi}^{2}-N_{\o \zeta}^2,\\
\pi_{\zeta}=N_{\zeta(\kappa \xi \o)}^{2}-N_{\zeta\kappa}^{2}-N_{\zeta\xi}^{2}-N_{\zeta \o}^2,
\end{eqnarray} from which we are able to obtain the $\pi_4$-tangle by
\begin{eqnarray}\label{eq28}
\pi_{4}=\frac{1}{4}\left(\pi_{\kappa}+\pi_{\xi}+\pi_{\o}+\pi_{\zeta}\right).
\end{eqnarray}

\begin{figure}[htbp]
\includegraphics[width=9cm]{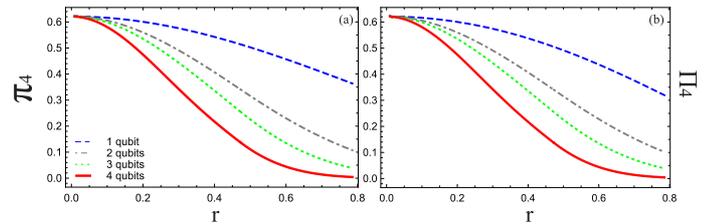}\label{pi-tangle}\hfil%
\caption{\label{Pi-tangle} (Color online) \text{Whole $\pi_{4}$-tangle} and geometric average $\Pi_4$ as a function of the acceleration parameter $r$ illustrated in (a) and (b) respectively. }
\end{figure}

Moreover, we might use another whole entanglement measurement defined as geometric mean \cite{Sabin} to describe the entanglement property of this tetrapartite system
\begin{eqnarray}\label{eq29}
\Pi_4=\left(\pi_{\kappa} \pi_{\xi} \pi_{\o} \pi_{\zeta}\right)^{\frac{1}{4}}.
\end{eqnarray}

Likewise, we are going to omit the analytical results due to the size of the polynomials and show their corresponding plots in FIG. 4. In FIG. 5 we show a comparison between $\pi_{4}$ and $\Pi_{4}$. Their difference implies that the entangled system becomes more robust only when one qubit, say Alice in our case, is accelerated but other observers are stationary. However, we notice that either $\pi_{4}$-tangle or $\Pi_4$ can be used to describe the entanglement property of this system since because of their small difference between them when three qubits (Bob, Charlie and David) or four qubits (Alice, Bob, Charlie and David) are moving in uniform acceleration.

\begin{figure}[htbp]
\includegraphics[width=7cm]{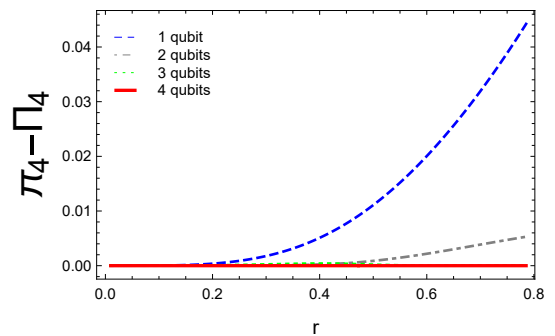}\label{pi-tangle}\hfil%
\caption{\label{Pi-tangle} \text{Difference between $\pi_{4}$ and $\Pi_{4}$} as a function of the acceleration parameter $r$. }
\end{figure}

\section{von Neumann entropy}

In order to know the measure of information  for an entangled quantum system it is necessary to study the von Neumann entropy defined as \cite{VoN},
\begin{eqnarray}
S=-\mathrm{Tr}(\rho \log_{2} \rho)=-\sum_{i=1}^{n} \lambda^{(i)} \log_{2}{\lambda^{(i)}}
\end{eqnarray}where $\lambda^{(i)}$ denotes the $i$-th eigenvalue of the density matrix $\rho$. It should be pointed out that the density matrix is not taken as its partial transpose. Based on this we are able to measure the degree of the satiability of the studied quantum state. We show the behaviour of the von Neumann entropy in FIG. 6. As expected, the von Neumann entropy of whole tetrapartite system increases with the increasing acceleration. It is more interesting to see that the von Neumann entropy becomes large with the number of accelerated observers as shown in panel (a) of FIG. 6.

\begin{figure}[htbp]
\subfigure[]{\includegraphics[width=7cm]{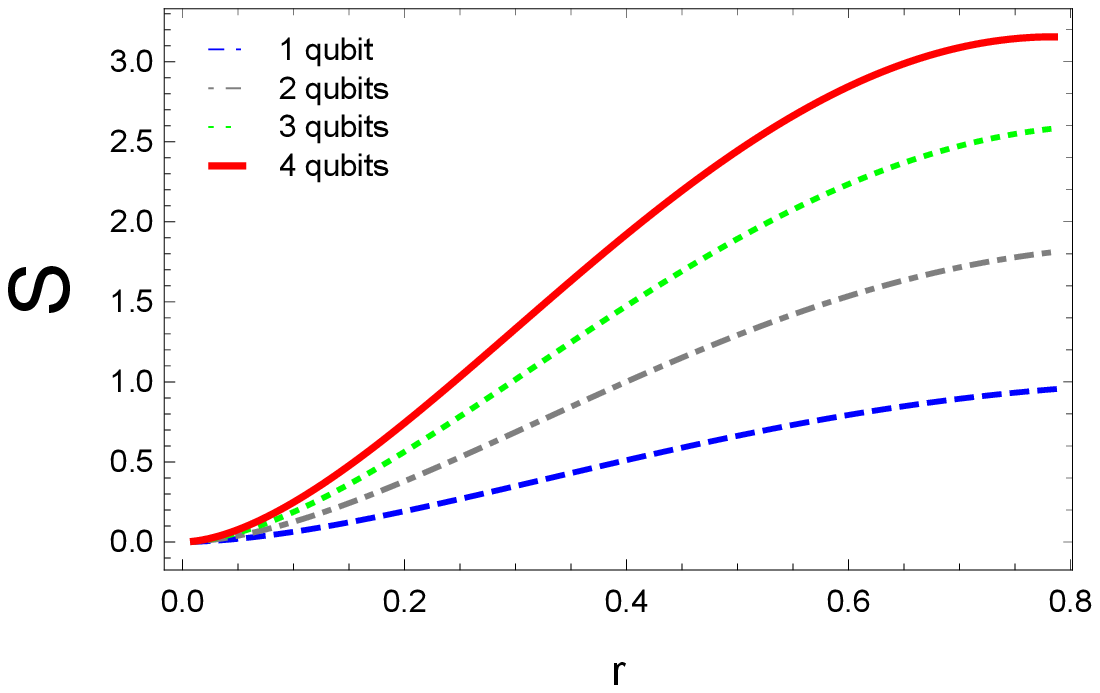}}\label{CA1}
\subfigure[]{\includegraphics[width=7cm]{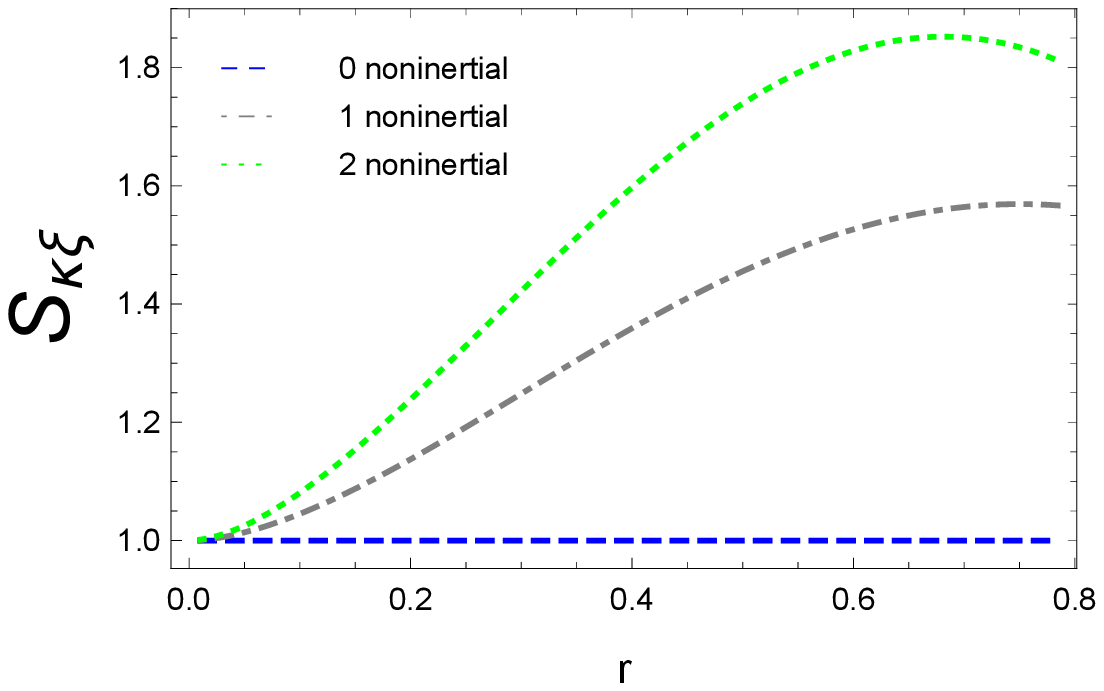}}\label{CA1}
\subfigure[]{\includegraphics[width=7cm]{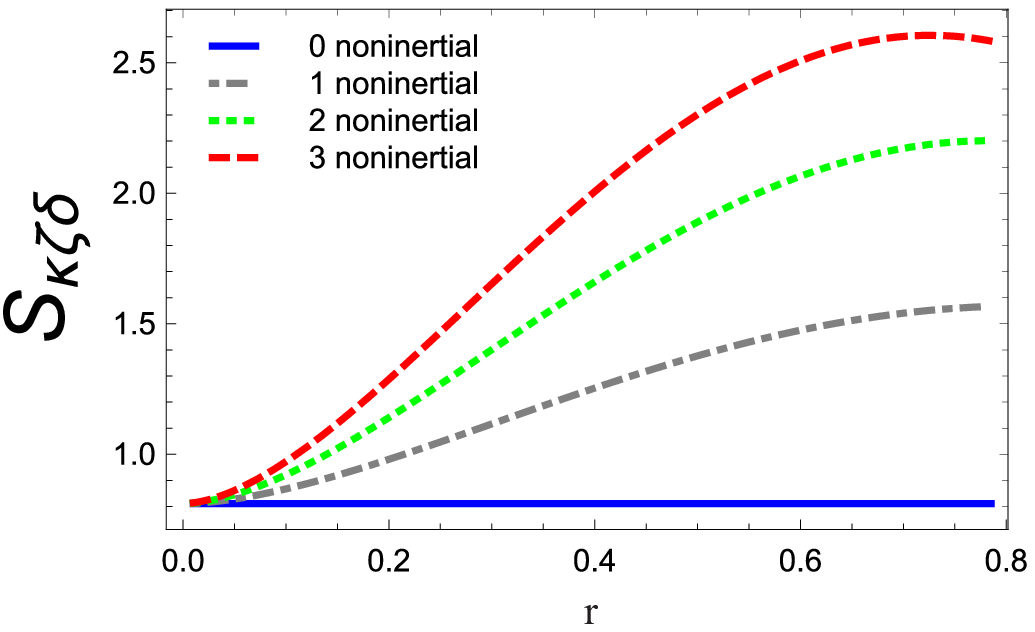}}\label{CA1}
\caption{ (color online) The Von Neumann Entropies $S$, $S_{\kappa\zeta\delta}$ and $S_{\kappa\xi}$ as a function of acceleration parameter $r$ are plotted in (a), (b) and (c) respectively. They correspond to the tetrapartite, tripartite and bipartite systems. }
\end{figure}

On the other hand, we show the subsystem entropies for the bipartite case, which exists only 3 possible entropy values. For the case when there is no any accelerated qubit, the entropy of the subsystem will be $S_{\kappa \xi}=1$. However, when the system has only one accelerated qubit we have the following eigenvalues:
\begin{equation}
\begin{array}{l}
\lambda_{\kappa_{I} \xi}^{(1)}=\frac{1}{2}\cos ^2 r, ~~~\lambda_{\kappa_{I} \xi}^{(2)}=\frac{1}{2}\sin ^2 r, \\[2mm]
\lambda_{\kappa_{I} \xi}^{(3, 4)}=\frac{1}{32} (10-2 \cos (2 r)\\[2mm]
~~~~~~~~\mp\sqrt{2} \sqrt{-20 \cos (2 r)+9 \cos (4 r)+43}),
\end{array}
\end{equation}where $"\mp"$ corresponds to $\lambda_{\kappa_{I} \xi}^{(3)}$ and $\lambda_{\kappa_{I} \xi}^{(4)}$, respectively. On the other hand, we find the eigenvalues for the  bipartite system which has all the qubits accelerated,
\begin{equation}
\begin{array}{l}
\lambda_{\kappa_{I} \xi_{I}}^{(1)}=\frac{1}{2}\cos ^4 r\\[2mm]
\lambda_{\kappa_{I} \xi_{I}}^{(2)}=\frac{1}{16} [1-\cos (4 r)], \\[2mm]
\lambda_{\kappa_{I} \xi_{I}}^{(3)}=\frac{1}{16} [4 \cos (2 r)-\cos (4 r)+5], \\[2mm]
\lambda_{\kappa_{I} \xi_{I}}^{(4)}=-\frac{1}{4} \sin ^2(r) [\cos (2 r)-3].
\end{array}
\end{equation} The von Neumann entropies for the cases when one and two observers are accelerated in uniform acceleration are illustrated in panel (b) of FIG. 6. Finally, let us consider the tripartite systems which include all possible combinations, e. g. without accelerated observer and with one, two and three accelerated observers. When there is no any accelerated observer, the eigenvalues are given by $3/4$ and $1/4$ so one has $S_{\kappa\zeta\delta}=0. 811278$. When the tripartite system has only one accelerated observer, the eigenvalues are given by
\begin{equation}\label{}
\begin{array}{l}
\lambda_{\kappa_{I} \zeta\delta}^{(1)}=\frac{1}{4}\cos ^2(r),\\[2mm]
\lambda_{\kappa_{I} \zeta\delta}^{(2)}=\frac{1}{4} [1-\cos (2 r)],\\[2mm]
\lambda_{\kappa_{I} \zeta\delta}^{(3,4)}=\frac{1}{32} [2 \cos (2 r)\\[2mm]
~~~~~\mp\sqrt{2} \sqrt{20 \cos (2 r)+9 \cos (4 r)+43}+10],
\end{array}
\end{equation}where the symbols $"\mp"$ correspond to $\lambda_{\kappa_{I} \zeta\delta}^{(3)}$ and $\lambda_{\kappa_{I} \zeta\delta}^{(4)}$, respectively.

When the tripartite system has two accelerated observers, the eigenvalues are given by
\begin{equation}\label{}
\begin{array}{l}
\lambda_{\kappa_{I} \zeta_{I}\delta}^{(1)}=\frac{1}{4}\cos ^4(r),\\[2mm]
\lambda_{\kappa_{I} \zeta_{I}\delta}^{(2)}=\lambda_{\kappa_{I} \zeta_{I}\delta}^{(2)}=\frac{1}{32} (1-\cos (4 r)),\\[2mm]
\lambda_{\kappa_{I} \zeta_{I}\delta}^{(4,5)}=\frac{1}{16} \Big[3 \cos (2 r)+3\\[2mm]
~~~~~~~\mp\sqrt{2} \sqrt{\cos (4 r) \cos ^4(r)+17 \cos ^4(r)}\Big],\\[2mm]
\lambda_{\kappa_{I} \zeta_{I}\delta}^{(6,7)}=3-3 \cos (2 r)\\[2mm]
~~~~~~~\mp\sqrt{2} \sqrt{17 \sin ^4(r)+\sin ^4(r) \cos (4 r)},
\end{array}
\end{equation}

When the tripartite system has three accelerated observers, the eigenvalues are given by
\begin{equation}\label{}
\begin{array}{l}
\lambda_{\kappa_{I} \zeta_{I}\delta_{I}}^{(1)}=\frac{1}{4}\cos ^6(r),\\[2mm]
\lambda_{\kappa_{I} \zeta_{I}\delta_{I}}^{(2)}=\lambda_{\kappa_{I} \zeta_{I}\delta}^{(3)}=\frac{1}{128} [\cos (2 r)-2 \cos (4 r)-\cos (6 r)+2],\\[2mm]
\lambda_{\kappa_{I} \zeta_{I}\delta_{I}}^{(4)}=\frac{1}{128} [49 \cos (2 r)+10 \cos (4 r)-\cos (6 r)+38],\\[2mm]
\lambda_{\kappa_{I} \zeta_{I}\delta_{I}}^{(5)}=\frac{1}{128} [-\cos (2 r)-18 \cos (4 r)+\cos (6 r)+18],\\[2mm]
\lambda_{\kappa_{I} \zeta_{I}\delta_{I}}^{(6,7)}=\frac{1}{128} [-\cos (2 r)-6 \cos (4 r)+\cos (6 r)+6],\\[2mm]
\lambda_{\kappa_{I} \zeta_{I}\delta_{I}}^{(8)}=-\frac{1}{8} \sin ^4(r) [\cos (2 r)-7].
\end{array}
\end{equation}
Their von Neumann entropies are shown in panel (c) of FIG. 6. The solid blue, dot-dashed grey, dotted green and dashed red lines represent the von Neumann entropies for the cases without any accelerated observer, with one, two and three accelerated observers, respectively.

\section{Discussions and concluding remarks}

In this work  we first computed the negativity of the entangled W-Class tetrapartite state. We have noticed that there exists disentanglement, i.e. entanglement of sudden death, for $1-1$ tangle case when $r>0. 472473$ only when four observers are accelerated at the same time. Other cases for $1-1$ tangle and those for $1-3$ tangle, however, are always entangled. On the other hand, we have reverified the fact that entanglement is an observer-dependent quantity in noninertial frame. When we compare the whole entanglement measures such as the arithmetic average $\pi_{4}$ and geometric average value $\Pi_{4}$, it is seen that for the cases when the system has one or two accelerated qubits there is a significant difference, that is, the arithmetic average value $\pi_{4}$ is greater than the geometric average value $\Pi_{4}$. However, when the system depends on three and four accelerated qubits, we find that their difference is almost zero. This implies that we might make use of either $\pi_{4}$ or $\Pi_{4}$ to describe this entangled system.

For the von Neumann entropy we have observed that the entropy increases as the number of accelerated qubits increases in the system. Moreover, we have noticed that the von Neumann entropies for both bipartite and tripartite  subsystems $S_{\kappa_{I}\xi_{I}}$ and $S_{\kappa_{I} \zeta_{I}\delta_{I}}$ are measured, we can see that they arrive to a maximum entropy and then begin to decrease. This implies that the subsystems $\rho_{\kappa_{I}\xi_{I}}$ and $\rho_{\kappa_{I} \zeta_{I}\delta_{I}}$ are first more disorder and then the disorder is reduced with the increasing acceleration. In addition, we note that the von Neumann entropies are calculated as $S_{\kappa\xi\delta\eta}=0$, $S_{\kappa\zeta\delta}=0. 811278$ and $S_{\kappa\xi}=1$, which correspond to the tetrapartite, tripartite and bipartite cases without any accelerated observer. This implies that
the system with more observers which are in stationary case is more stable. Before ending this work, we give a useful remark on the acceleration limit value $r$. As we know, there exists the disentanglement phenomenon after the acceleration $r \approx 0. 472473$ only when the $1-1$ tangle of the W-Class tetrapartite includes two accelerated observers. This value is different from that of GHZ tetrapartite system, in which this value was given by $r\approx 0. 417$. This implies that the present W-Class tetrapartite system is more robust than that of the GHZ tetrapartite system since $r_{\rm W-Class} \approx 0. 472473> r_{\rm GHZ} \approx 0. 417$. {\color{red}Finally, it should be mentioned that we are going to see whether or how the present study extends to the thermodynamic properties as treated in \cite{hass1, hass2}.}

\section*{Acknowledgements}
We thank the referees for making invaluable suggestions. This work was partially supported by the CONACYT, Mexico under the Grant No. 288856-CB-2016 and partially by 20190234-SIP-IPN, Mexico.

\end{document}